\documentclass[hidelinks, review]{elsarticle}

\usepackage{amssymb}

\usepackage{tablefootnote}
\usepackage{graphicx}
\usepackage{lineno,hyperref}
\usepackage{tabularx}
\usepackage{chemformula}
\usepackage{hhline} 
\usepackage{multirow} 
\usepackage{subfig}
\usepackage{float}
\usepackage{siunitx}
\usepackage{mathtools}
\usepackage{float}
\restylefloat{table}

\setcitestyle{square}
\usepackage{caption} 
\usepackage{threeparttable}
\modulolinenumbers[5]
\journal{Materials Today Commmunications}
\begin{document}

\begin{frontmatter}



\title{Probing the interactions between interstitial hydrogen atoms in niobium through density functional theory calculations}


\author{Arvind Ramachandran*}
\ead{aramac13@asu.edu}
\address{School of Sustainable Engineering and the Built Environment, Arizona State University, Tempe, AZ 85281, USA}

\author{Houlong Zhuang}
\ead{hzhuang7@asu.edu }
\address{School for Engineering of Matter, Transport and Energy, Arizona State University, Tempe, AZ 85287, USA}

\author{Klaus S. Lackner}
\ead{Klaus.Lackner@asu.edu}
\address{School of Sustainable Engineering and the Built Environment, Arizona State University, Tempe, AZ 85281, USA}

\cortext[mycorrespondingauthor]{Corresponding author}

\begin{abstract}

Past experiments about hydrogen absorption in niobium have revealed specific properties about interactions between interstitial hydrogen atoms. It has been reported that there are long-range attractive and short-range repulsive interactions between interstitial hydrogen atoms in niobium. It has also been reported that these interactions are of many-body nature. While previous understanding of these interactions is based on experimental inferences from past experiments, through these calculations, for the first time, we can understand the nature of the interactions at a fundamental level. In this work, we use Density Functional Theory calculations to study the interactions of interstitial hydrogen atoms in niobium. We report here that these interactions are a combination of an attractive, indirect image interaction and a repulsive, direct interaction. Through our calculations, we also infer here that these interactions indeed have many-body characteristics.

\end{abstract}

\begin{keyword}
Density functional theory, interstitials, alloys
\end{keyword}

\end{frontmatter}

\section{Introduction}
\label{intro}
The interactions of interstitial hydrogen atoms in group V transition metals such as niobium, vanadium, and tantalum is an experimentally well-studied problem. The three most unusual properties demonstrated experimentally in these metals are the high hydrogen absorption capacity \cite{veleckisThermodynamicPropertiesSystems1969a,albrechtReactionsNiobiumHydrogen1959a,kujiKuji1984Thermodynamic1984,komjathyNiobiumhydrogenSystem1960}, the facile thermal diffusion of hydrogen \cite{schaumanng.Schaumann1970Nb2006,wipfDiffusionCoefficientHeat}, and the electromigration of hydrogen that is characteristic of an unscreened proton \cite{petersonPeterson1978Nb1978,erckmannErckman1976Nb1976,brouwerBrouwer1989Nb1989,wipfWipf1975Metals1976}. \\

Experiments show that group V metals in their alpha phase can absorb a large amount of hydrogen \cite{veleckisThermodynamicPropertiesSystems1969a,albrechtReactionsNiobiumHydrogen1959a,kujiKuji1984Thermodynamic1984,komjathyNiobiumhydrogenSystem1960}. Hydrogen molecules from the gas phase dissociate, and the hydrogen atoms occupy interstitial positions in the metal lattice. The affinity trends of hydrogen in these metals are usually presented in the form of absorption isotherms. At low concentrations, following Sievert's law, the pressure of the equilibrated hydrogen gas varies as the square of the hydrogen concentration. Such low concentrations represent the ideal limit where the interstitial hydrogen atoms are non-interacting. However, this dependence does not hold at higher concentrations, indicating that hydrogen-hydrogen interactions start to become critical, and deviation from ideality occurs. These absorption isotherms reveal particular properties regarding the nature of the interaction between the interstitial hydrogen atoms in niobium. One of the inferences made is that there are long-range attractive and short-range repulsive interactions between the interstitial hydrogen atoms \cite{kujiKuji1984Thermodynamic1984,fukaiMetalHydrogenSystemBasic2005}. Another inference is that the interaction between the hydrogen atoms cannot be thought of as a two-body interaction and is best represented as a many-body interaction \cite{fukaiMetalHydrogenSystemBasic2005}. \\

Group V metals provide an environment for the facile thermal diffusion of interstitial hydrogen and deuterium. The diffusion constants for hydrogen and deuterium are very high ($~10^{-4}$ cm$^2$ s$^{-1}$). They are orders of magnitude larger than diffusion constants for other interstitial atoms like oxygen and nitrogen. Experiments by Schaumann et al.\ \cite{schaumanng.Schaumann1970Nb2006}, and Wipf et al.\ \cite{wipfDiffusionCoefficientHeat} reveal that there is a break in the apparent activation energy of the diffusion coefficient of hydrogen, which is best explained by a transition from an Arrhenius type behavior to quantum mechanical tunneling through potential barriers. These experimenters observed that this break does not occur for deuterium diffusion, which can be explained by the higher mass of deuterium, and thus a lower tunneling probability. The experimental diffusion constant, its dependence on temperature, and the observed isotope effects are inconsistent with the classical theory of diffusion \cite{schaumanng.Schaumann1970Nb2006}. \\

Finally, experiments have shown that hydrogen inside the structure of these metals responds to weak electric fields \cite{petersonPeterson1978Nb1978,erckmannErckman1976Nb1976,brouwerBrouwer1989Nb1989,wipfWipf1975Metals1976}. If there is an electric current flowing through the metal, hydrogen in the metal flows in the opposite direction to electrons, suggesting that the hydrogen inside the metal has a positive effective charge that responds to a very small gradient overlaid on the periodic potential from the crystal structure in which it can move essentially freely. However, electrons are expected to flow from these metal to hydrogen since hydrogen's electronegativity is larger than that of all three metals. The apparent contradiction hints that the electromigration of hydrogen in these metals is more involved than a simple model of a free charged particle accelerating in the presence of an electric field. \\

The focus of this paper will be on the interactions of hydrogen in niobium. Therefore, we begin by summarizing the understanding of these interactions from experiments. As mentioned earlier, the affinity of hydrogen to metals is presented in the form of absorption isotherms \cite{veleckisThermodynamicPropertiesSystems1969a,albrechtReactionsNiobiumHydrogen1959a,kujiKuji1984Thermodynamic1984,komjathyNiobiumhydrogenSystem1960}. The absorption isotherms are a direct measure of the Gibbs free energy of hydrogen dissolution in niobium. From the Gibbs free energy, one can deduce the enthalpy and entropy of hydrogen dissolution in niobium, both of which reveal important properties about the interactions of interstitial hydrogen atoms in niobium. \\

At low concentrations, the enthalpy of hydrogen dissolution in niobium ($\Delta H$) decreases with concentration, and at higher concentrations, this trend starts to reverse. The concentration of hydrogen in these metal-hydrogen systems is usually expressed as the hydrogen to metal ratio ($\mathrm{\frac{H}{M}}$). $\Delta H$ decreases from its value at infinite dilution of -0.35 eV/H atom to -0.46 eV/H atom at $\mathrm{\frac{H}{Nb}} = 0.57$. For $\mathrm{\frac{H}{Nb}} > 0.57$, $\Delta H$ begins to increase with increasing $\mathrm{\frac{H}{Nb}}$ values, attaining a value of -0.34 eV/H atom at $\mathrm{\frac{H}{Nb}} = 0.83$ \cite{kujiKuji1984Thermodynamic1984}. \\

Alefeld \cite{alefeldAlefeld1972Phase1972} noted that an important contribution to the enthalpy of hydrogen dissolution in metals comes from elastic interaction. A hydrogen atom when occupying an interstitial site displaces the neighboring atoms, which in turn displace the atoms in the next neighboring shell and so on \cite{alefeldAlefeld1972Phase1972}. This type of interaction leads to a strain field in the lattice, increasing the volume occupied by the niobium lattice by 2-3 $\mathrm{\si{\angstrom}^3}$ per H atom \cite{fukaiMetalHydrogenSystemBasic2005}. This additional volume is effectively the volume occupied by a hydrogen atom. For comparison, the volume of a niobium atom occupies 17.97 $\mathrm{\si{\angstrom}^3}$ \cite{kittel2005introduction}. The contribution of this elastic interaction to the $\Delta H$ is called the volume interaction energy or the indirect image interaction energy \cite{kujiKuji1984Thermodynamic1984}. \\

In the presence of multiple hydrogen atoms, one can think of the interaction of a hydrogen atom with the strain field caused by other hydrogen atoms. Alefeld \cite{alefeldAlefeld1972Phase1972} considered this type of mean-field treatment of this interaction and showed that for density changes in which the wavelengths are comparable to the sample size, the elastic interaction is attractive and is responsible for the observed decrease in $\Delta H$ at low concentrations. The strain field created by a hydrogen atom falls off as the inverse square of the distance from that hydrogen atom, and hence the interaction of two hydrogen atoms via their strain fields is of long-range order. Alefeld \cite{alefeldAlefeld1972Phase1972} called this interaction between two hydrogen atoms via their long-range strain fields as the ``elastic dipole-dipole interaction." \\

Conceptually, this interaction is no different from the indirect image interaction described earlier, as they are both related to the elastic interaction caused by hydrogen atoms in niobium. \\

Another contribution to the enthalpy of dissolution comes from the direct interaction between hydrogen atoms in the niobium lattice. This direct interaction between the hydrogen atoms in niobium is unlike the indirect image interaction energy in that it is repulsive, is of short-range order, and depends on the exact configuration of hydrogen atoms. At low concentrations, the attractive indirect image interaction dominates, but at higher concentrations, the repulsive direct interaction starts to dominate $\Delta H$ \cite{kujiKuji1984Thermodynamic1984,fukaiMetalHydrogenSystemBasic2005}. \\ 

Kuji et al.\ \cite{kujiKuji1984Thermodynamic1984} noted that even at the highest concentrations of hydrogen observed, only a small fraction of the total number of sites available is occupied. Their reasoning behind this observation is that there is a mutual occupancy blocking between hydrogen atoms, i.e., occupation of a site by a hydrogen atom precludes the occupation of sites in its vicinity \cite{kujiKuji1984Thermodynamic1984}. \\ 

The experimentally observed trends for the entropy of hydrogen dissolution in niobium ($\Delta S$) further supports this reasoning. Kuji et al.\ \cite{kujiKuji1984Thermodynamic1984} compared the experimental $\Delta S$ with several configurational entropy model calculations. They found that the model that assumes a mutual blocking of the first two nearest neighboring sites agrees closest to the experimental values. This observation led them to conclude that the repulsive interaction energies between pairs of hydrogen atoms closer than the second nearest neighboring distance are so large that configurations with hydrogen atoms closer than this distance are not physically realized. Further, Kuji et al.\ \cite{kujiKuji1984Thermodynamic1984} also noted that for $\mathrm{\frac{H}{Nb}}<0.75$, neighbors closer than the seventh nearest neighbor could be avoided. Since there are substantial repulsive contributions to $\Delta H$ starting from $\mathrm{\frac{H}{Nb}} = 0.5$, they concluded that in addition to the very strong repulsion observed within the second neighboring sites, there have to be weaker, although significant repulsive interactions between hydrogen atoms that are farther apart than the second neighbor distance. \\

Fukai \cite{fukaiMetalHydrogenSystemBasic2005} quotes the Westlake criteria, which is an empirical rule that states interstitial hydrogen atoms do not come closer than 2.1 $\si{\angstrom}$ in metal-hydrogen systems, as another indicator of strong short-range repulsive interactions between hydrogen atoms. Finally, Fukai \cite{fukaiMetalHydrogenSystemBasic2005} also argues that the formation of ordered structures in metal-hydrogen systems is another indicator of the short-order repulsion. If the interaction between the interstitial hydrogen atoms were purely attractive, we would observe precipitation of hydrogen-rich phases at low temperatures instead of the formation of ordered structures. \\

\begin{figure}[H]
\centering
  \includegraphics[width=\textwidth]{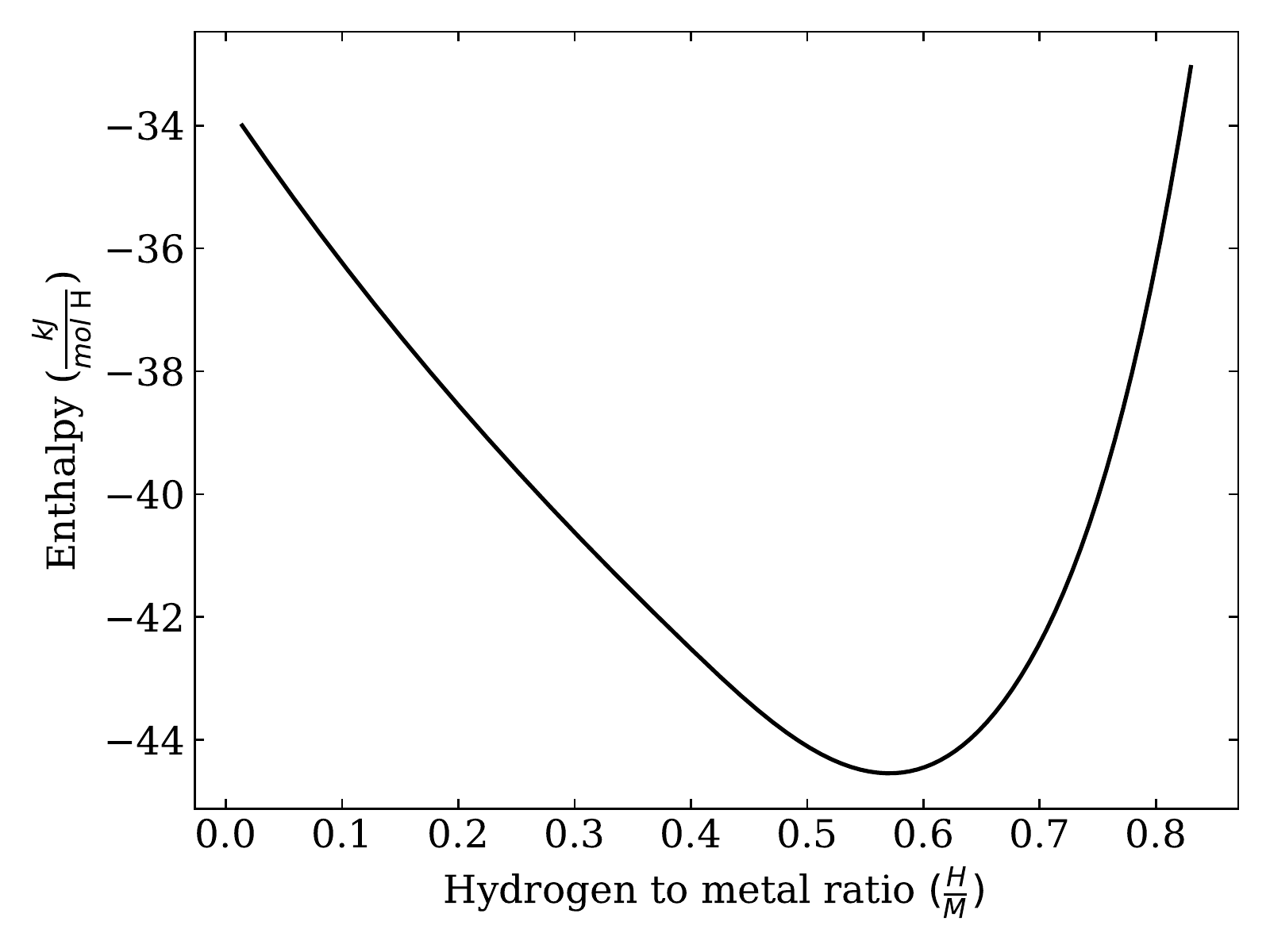}
\caption{Enthalpy of hydrogen dissolution in niobium \cite{kujiKuji1984Thermodynamic1984}.}
\label{fig:1}       
\end{figure}

In Fig.\ \ref{fig:1}, we plot the variation of the enthalpy of dissolution with concentration, obtained from Kuji et al.\ \cite{kujiKuji1984Thermodynamic1984}. Fig.\ \ref{fig:1} shows that the variation of $\Delta H$ with hydrogen concentration is not linear. Fukai \cite{fukaiMetalHydrogenSystemBasic2005} argues that the nonlinear variation of $\Delta H$ with hydrogen concentration indicates that the interactions between hydrogen atoms are not strictly pairwise. Because if it were, the extensive enthalpy of dissolution would be quadratic in hydrogen concentration, and hence $\Delta H$, would be linear. Therefore, the interactions between the hydrogen atoms are best represented as a many-body problem. Oates and Stoneham.\ \cite{Oates_1983} noted that in Pd, the energies of clusters of hydrogen atoms could not be expressed as a sum of pairwise interactions, or in other words, many-body effects are present. \\

The current understanding of these interactions is based on inferences from experimental results. A consistent framework that explains all of these properties, starting from first principles, is lacking. Our high-level goal is to use first principle methods to explain these properties, and this work is our first effort toward this goal. The approach we will take is to start with the niobium-hydrogen system and work our way to other systems. The scope of this paper is to explain some of the observations made about the interactions between the interstitial hydrogen atoms in niobium through density functional theory calculations. \\

We begin by calculating fundamental quantities of niobium, such as the lattice constant, cohesive energy (defined in Eq.\ \ref{eq:1}), and elastic constants ($C_{11}$, $C_{12}$, $C_{44}$). We progress to studying hydrogen in niobium by identifying the preferred site occupancy of a single hydrogen atom in niobium supercells of varying sizes. We then study the two-hydrogen problem very rigorously considering all possible configurations of the two-hydrogen system. Finally, we present some results on the many-body nature of the interaction of interstitial hydrogen atoms in niobium.

\section{Computational Methods}
\label{methods}
We perform density functional theory (DFT) simulations using the plane-wave code Vienna ab initio simulation package (VASP, version 5.4.4). We used the Projector-augmented wave method and deployed the Perdew-Burke-Ernzerhof (PBE) functional for all calculations \cite{Ref9,Ref10,Ref11,Ref16}. The cutoff energy for the plane wave basis set is set to 600~eV to ensure an energy accuracy of 1~meV/atom. For the $\mathrm{k}$-point sampling, we used the Monkhorst-Pack scheme \cite{Ref19}. We employed a $24\times 24\times 24$ mesh for the structural relaxations of the two-atom BCC niobium unit cell. For all other calculations, we scaled the $\mathrm{k}$-points relative to the size of the supercell. For all calculations involving cell shape and volume relaxation, succesive re-relaxation calculations were performed until structural convergence was observed. Zero-point energy corrections are not included in the calculations because we are only interested in studying various energy trends in this paper. However, we will consider these corrections in future work, where we intend to compare such trends with experimental data.

\section{Results and Discussion} 

We calculated fundamental quantities such as the lattice constant, cohesive energy (defined in Eq.\ \ref{eq:1}), and elastic constants ($C_{11}$, $C_{12}$, $C_{44}$). In Eq.\ \ref{eq:1}, $E_\mathrm{atom}$ is the energy of an isolated Nb atom in a rectangular vacuum box. The half factor comes from having two atoms in the unit cell of Nb. We compare the results from these calculations with experimental values in Table \ref{tab:1}. The predicted values seem to be in good agreement with the experimental values of these quantities. Therefore, we used the PBE functional in all subsequent calculations reported in this paper. 

\begin{equation}
  \label{eq:1}
E_\mathrm{coh} = E_\mathrm{atom} - E_\mathrm{bulk}/2 
\end{equation}

\begin{table}[h]
  \normalsize
  \begin{threeparttable}[b]
  \caption{Comparison of the predicted lattice constant $a_0$  ($\si{\angstrom}$), cohesive energy $E_\mathrm{coh}$(eV/atom), and elastic constants $C_{11}$, $C_{12}$, and $C_{44}$ (GPa) of Nb using the PBE functional with experimental values.}
     \begin{tabular}{ccccccc}

      & $a_0$ & $E_\mathrm{coh}$ & $C_{11}$ & $C_{12}$  & $C_{44}$\\ 

      & ($\si{\angstrom}$) & (eV/atom) & (GPa) & (GPa) & (GPa) \\
       \hline 
       \\
      PBE &3.31 &7.01 &251$^\mathrm{1}$, 246$^\mathrm{2}$ & 132$^\mathrm{1}$, 133$^\mathrm{2}$  & 21$^\mathrm{1}$, 17$^\mathrm{2}$\\ \\
        Experiment &3.30$^\mathrm{3}$ &7.57$^\mathrm{3}$ & 253$^\mathrm{4}$ & 132$^\mathrm{4}$ & 31$^\mathrm{4}$\\ \\

      \end{tabular}

    \begin{tablenotes}{\footnotesize
      \item [1] Strain-energy method
      \item [2] Symmetry-general approach \cite{lepageSymmetrygeneralLeastsquaresExtraction2002}
      \item [3] Kittel \cite{kittel2005introduction}
      \item [4] Hayes et al. \cite{hayes1974elastic}
        }
    \end{tablenotes}
 \label{tab:1}
  \end{threeparttable}
\end{table}

In order to better understand the origin of the hydrogen-hydrogen interaction in metals, we present here the formalism for the enthalpy of hydrogen dissolution developed by Fukai \cite{fukaiMetalHydrogenSystemBasic2005}. One can express the dependence of the enthalpy of hydrogen dissolution on hydrogen concentration as 
\begin{equation}
  \label{eq:2}
\frac{\partial \Delta H}{\partial x}  = \left ( \frac{\partial \Delta H}{\partial V} \right )_x  \frac{\partial V}{\partial x}+ \left ( \frac{\partial \Delta H}{\partial x} \right )_V 
\end{equation}
The first term in Eq.\ \ref{eq:2} depends on the rate of change of volume with hydrogen concentration and represents the volume interaction energy or the indirect image interaction energy. The second term in Eq.\ \ref{eq:2}, which is the change of enthalpy at constant volume, represents the direct interaction between the hydrogen atoms. This viewpoint of separating the contributions to the enthalpy of hydrogen dissolution into the two different kinds of interactions gives us the ability to design calculations that study the origin of these interactions. \\

\begin{figure}[H]

\centering
 
\subfloat[H atom occupying a tetrahedral site (0,$\frac{1}{2}$,$\frac{1}{4}$)]{
  \label{Tetrahedral site}
  \includegraphics[width=8cm]{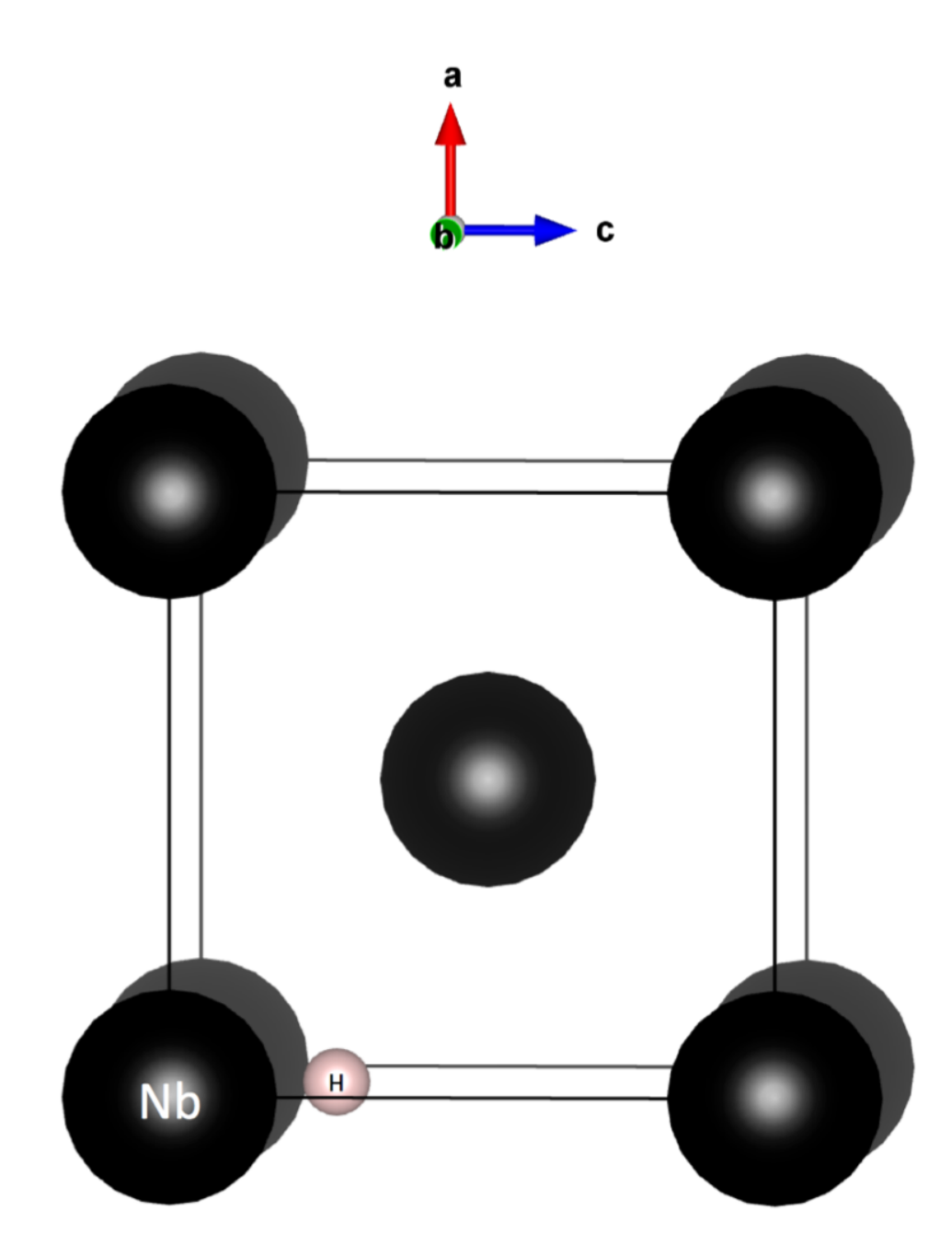} } 
 
\subfloat[H atom occupying a octahedral site (0,$\frac{1}{2}$,$\frac{1}{2}$)]{
  \label{Octahedral site}
  \includegraphics[width=8cm]{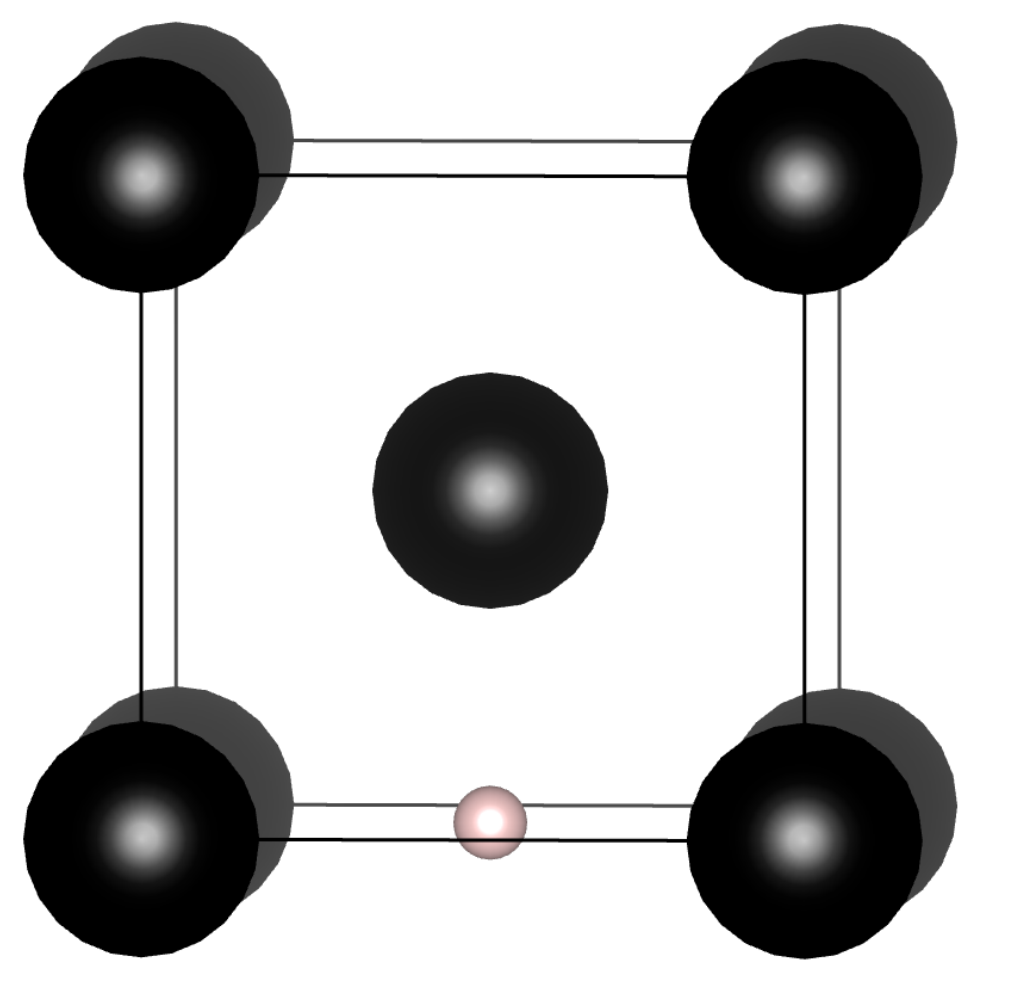}} 
 
\caption{Illustrations of H atoms occupying different sites in a Nb unit cell}
\label{fig:2}

\end{figure}

In the body-centered cubic (BCC) unit cell of niobium, interstitial hydrogen can occupy two different types of sites - the tetrahedral site and the octahedral site. We illustrate hydrogen atoms occupying the tetrahedral and octahedral sites in a niobium unit cell in Fig.\ \ref{fig:2}. \\

In Eq.\ \ref{eq:3}, we define the absorption energy of a single hydrogen atom in a niobium supercell. 

\begin{equation}
  \label{eq:3}
E_\mathrm{abs} = E_\mathrm{supercell+H}-E_\mathrm{supercell}-\frac{1}{2}E_\mathrm{H_2}
\end{equation}

$E_\mathrm{H_2}$ in Eq. \ref{eq:3} was calculated using DFT by placing an isolated hydrogen molecule in the center of a supercell of edge 10 $\si{\angstrom}$ and allowing for the H ion positions to relax. Using Eq. \ref{eq:3}, we calculated the absorption energies of a single hydrogen atom occupying tetrahedral and octahedral sites in a $1\times 1\times 1$, $2\times 2\times 2$, $3\times 3\times 3$, and $4\times 4\times 4$ supercells. The niobium and hydrogen ion positions, cell shape, and cell volume were relaxed for the relaxed calculations. Only the niobium and hydrogen ion positions were relaxed for the unrelaxed calculations. The difference in volume between the relaxed and the unrelaxed cells per hydrogen atom is $\nu_H$. The difference in absorption energies in the relaxed and the unrelaxed cell is the relaxation energy or $E_\mathrm{relaxation}$ \\

We report the $E_\mathrm{abs}$, $E_\mathrm{relaxation}$, and the $\nu_H$ values for the various calculations in Table \ref{tab:2}. \\

 \begin{table}[H]
\normalsize
  \caption{$E_\mathrm{abs}$, $E_\mathrm{relaxation}$, and the $\nu_H$ values for the various supercells studied. The superscripts t and o refer to the tetrahedral and octahedral sites, respectively.}
    \begin{center}
      \begin{tabular}{ccccc}

        Supercell & \multicolumn{2}{c}{$E_\mathrm{abs}$ (eV)} & $E_\mathrm{relaxation}$ (eV) &$\nu_H$ ($\si{\angstrom}^3$)\\  

        & Unrelaxed & Relaxed & \\  

        \hline \\

        \multirow{2}{*}{$1\times1\times1$} &-0.28$^\mathrm{t}$ &-0.43$^\mathrm{t}$ & -0.15$^\mathrm{t}$ & 3.07$^\mathrm{t}$\\
        &0.38$^\mathrm{o}$ &-0.22$^\mathrm{o}$ & -0.60$^\mathrm{0}$ & 2.58$^\mathrm{o}$ \\ \\
        \multirow{2}{*}{$2\times2\times2$} &-0.37$^\mathrm{t}$ &-0.39$^\mathrm{t}$ & -0.02$^\mathrm{t}$ & 3.09$^\mathrm{t}$\\
        &0.00$^\mathrm{o}$&-0.08$^\mathrm{o}$ & -0.08$^\mathrm{0}$ & 3.18$^\mathrm{o}$  \\ \\
        \multirow{2}{*}{$3\times3\times3$} &-0.41$^\mathrm{t}$ &-0.41$^\mathrm{t}$ & \  0.00 $^\mathrm{t}$ & 3.03$^\mathrm{t}$ \\
        &-0.12$^\mathrm{o}$&-0.14$^\mathrm{o}$ & -0.02$^\mathrm{0}$ & 3.06$^\mathrm{o}$  \\ \\
        \multirow{2}{*}{$4\times4\times4$} &-0.41$^\mathrm{t}$ &-0.41$^\mathrm{t}$ & \  0.00 $^\mathrm{t}$ & 3.27$^\mathrm{t}$\\
        &-0.14$^\mathrm{o}$&-0.15$^\mathrm{o}$ & -0.01$^\mathrm{0}$ & 2.43$^\mathrm{o}$\\ \\
      \end{tabular}
    \end{center}
  \label{tab:2}
\end{table}

Studying the difference between the energies of the relaxed and the unrelaxed calculations is a clear way of measuring the first term in Eq.\ \ref{eq:2}, the volume interaction energy caused by the hydrogen atom in the niobium supercell. \\

From Table \ref{tab:2}, we make the following observations. Firstly, interstitial hydrogen in niobium prefers to occupy the tetrahedral site at the different concentrations explored, regardless of whether or not the calculations were relaxed. Further, for all the supercells, the relaxed absorption energies are lower than or equal to the unrelaxed absorption energies. This result is clear evidence that the volume interaction of the hydrogen atom is attractive. The relaxation energy, $E_\mathrm{relaxation}$, becomes smaller in magnitude as the supercell size becomes larger, which makes sense because with $\nu_H$ being roughly constant, the volume strain becomes smaller as the supercell size grows.  Finally, we note that the $\nu_H$ values calculated for tetrahedral occupation in all the supercells are in close agreement with low concentration experimental value of 3.13 $\si{\angstrom}^3$ \cite{fukaiMetalHydrogenSystemBasic2005}. \\

As discussed before, we know that the hydrogen prefers to occupy the tetrahedral site in bulk niobium. The first hydrogen atom in the supercell can thus be populated to any of the tetrahedral sites in the bulk lattice, as they are all symmetrically equivalent. However, the second hydrogen atom in the supercell can be at different tetrahedral sites relative to the first hydrogen atom. To systematically study this problem, we considered a $3\times 3\times 3$ niobium supercell with two hydrogen atoms in different tetrahedral sites and constructed configurations with all possible interatomic distances between the hydrogen atoms. We only considered one configuration for every possible interatomic distance, as we treated configurations with the same interatomic distance as symmetrically equivalent. We consider a maximum interatomic distance equal to half the length of the body diagonal of the supercell. Any pairs that appear further apart are indeed closer because of the inherent periodicity constraint of the calculations. It turns out there are 36 possible unique distances, and we, therefore, considered 36 different configurations. We calculated the energies of these configurations using DFT, allowing for the ion positions, cell shape, and the cell volume to be relaxed. \\

\begin{figure}[H]
\centering
 \includegraphics[width=\textwidth]{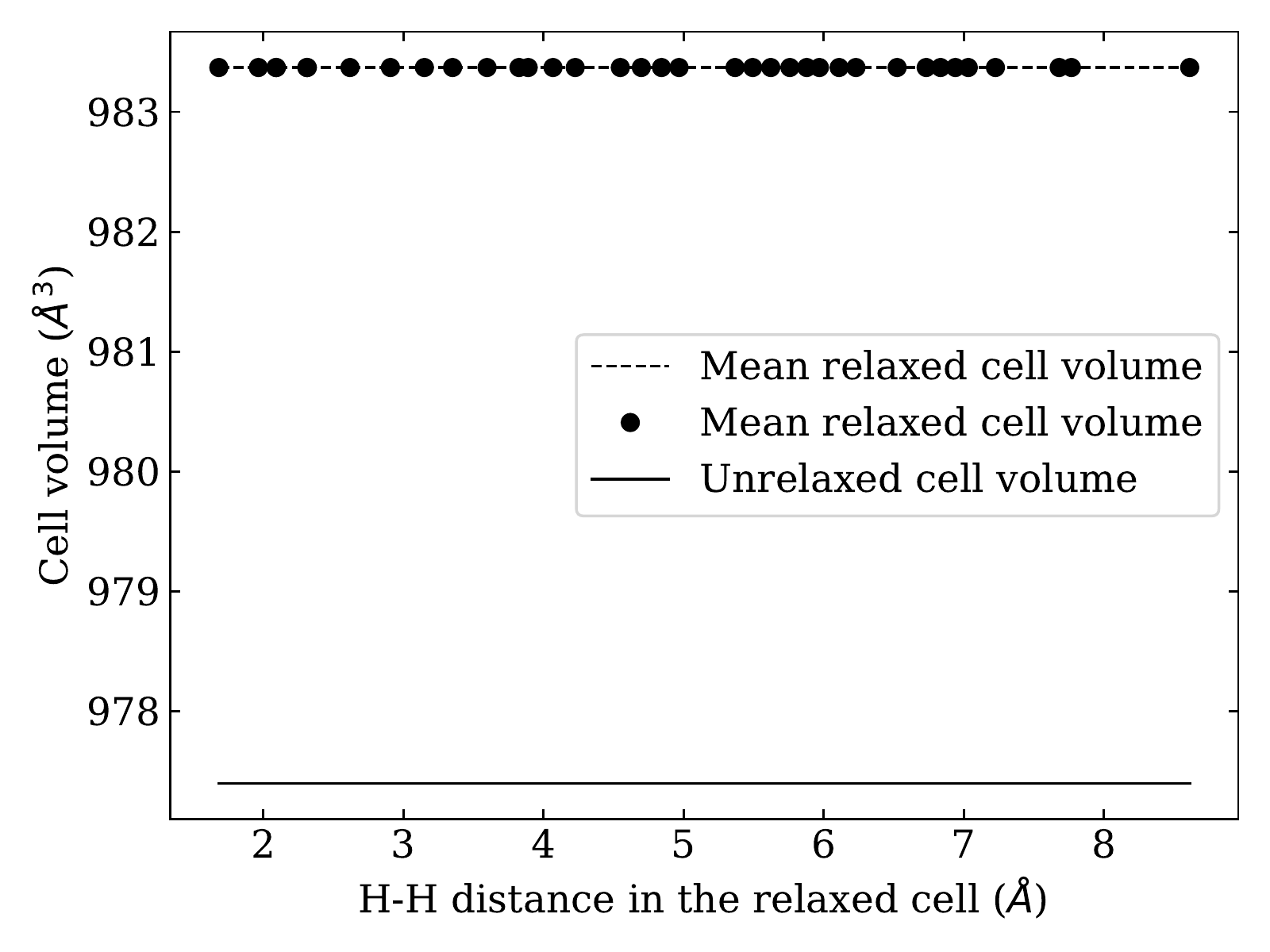}
 \caption{Relaxed and unrelaxed cell volumes of the different configurations of the two-hydrogen system.}
 \label{fig:3}
\end{figure}

The mean value of the cell volume of the different configurations is 983.37 $\si{\angstrom}^3$, and the standard deviation is 0.10 $\si{\angstrom}^3$. The mean value is 5.97 $\si{\angstrom}^3$ higher than the cell volume of the niobium cell with no interstitial hydrogen. In other words, the cell volume increase per H atom, $\nu_H$, is 2.99 $\si{\angstrom}^3$. We present these cell volumes in Fig.\ \ref{fig:3}. As can be seen visually and through the standard deviation metric, the relaxed cell volumes of every configuration are nearly the same. In Eq.\ \ref{eq:2}, since the first term vanishes at constant volume, we infer that the difference in the energies of the different configurations is a result of the differences in the direct interaction energies. To compare the energies of the different configurations, we reference the energies of the different configurations as in Eq.\ \ref{eq:4} to obtain the interaction energies per hydrogen atom. \\

\begin{equation}
 \label{eq:4}
E_\mathrm{interaction} =  E_\mathrm{supercell+2H}-2E_\mathrm{supercell+H}+E_\mathrm{supercell}
\end{equation}

We note that the $\nu_H$ for a two-hydrogen system in a $3\times 3\times 3$ niobium supercell (2.98 $\si{\angstrom}^3$), is slightly smaller than the value for the single hydrogen system in a $3\times 3\times 3$ niobium supercell (3.03 $\si{\angstrom}^3$ refer Table \ref{tab:2}). So the interaction energy in Eq.\ \ref{eq:4} is not an exact measure of the direct interaction between the hydrogen atoms. The two hydrogen atoms in the two-hydrogen system pack closer together than two non-interacting hydrogen atoms would, indicating that the elastic interaction between the two hydrogen atoms, i.e., the indirect image interaction, is attractive. This observation is consistent with the nature of the elastic dipole-dipole interaction described by Alefeld \cite{alefeldAlefeld1972Phase1972}. \\

\begin{figure}[H]
\centering
 \includegraphics[width=\textwidth]{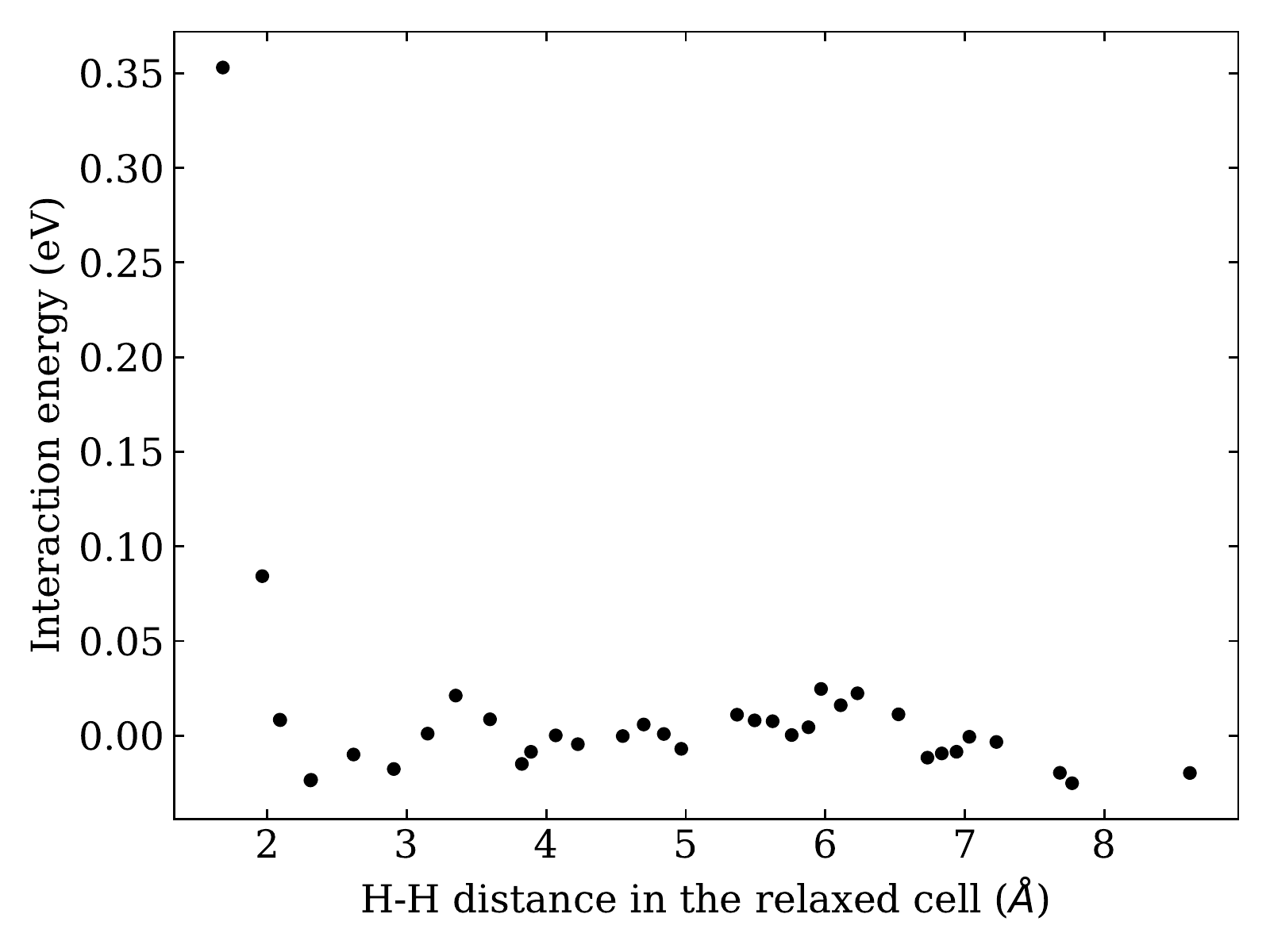}
 \caption{Interaction energies of the different configurations of the two-hydrogen system.}
 \label{fig:4}
\end{figure}

Nevertheless, since the $\nu_H$ values are very similar for the single-hydrogen and two-hydrogen systems, there is very little strain-field related interaction between the two hydrogen atoms in the two-hydrogen system. Therefore, we infer that the interaction energy we calculated is a good measure of the direct interaction between the hydrogen atoms in niobium. We plot the interaction energies as a function of the interatomic distance in 
Fig.\ \ref{fig:4}. The interaction energy depends on how far apart the hydrogen atoms are, indicating that that the direct interaction energy depends on the exact configuration of the hydrogen atoms. The interaction energies between hydrogen atoms closer than the second nearest neighbor are significant, while the interaction energies between hydrogen atoms that are farther apart are pretty close to zero. This explains why Kuji et al. \cite{kujiKuji1984Thermodynamic1984} found that the configurational entropy model that assumes a mutual blocking of the first two nearest neighboring sites agrees closest to the experimental values. We also find that for configurations with pairwise hydrogen distance greater than 2.1 $\si{\angstrom}$, the interaction energies are pretty close to zero, validating the Westlake criteria for metal-hydrogen systems.

To demonstrate the many-body nature of the hydrogen interaction in niobium, we will compare the interaction energies of several configurations of many-hydrogen systems calculated using DFT, with the respective interaction energies predicted from single-particle and pairwise contributions. \\

The interaction energy of three hydrogen atoms with each other and the niobium structure is given by Eq.\ \ref{eq:5}

\begin{equation}
 \label{eq:5}
E_\mathrm{interaction} = E_\mathrm{supercell+nH}-E_\mathrm{supercell}-\frac{n}{2}E_\mathrm{H_2}
\end{equation}

The interaction can also be expanded according to the number of particles involved as in Eq.\ \ref{eq:6}. It includes the single hydrogen atom binding energy to the niobium structure, $E_\mathrm{i}$ (as in Eq.\ \ref{eq:3}), the pairwise interaction energies, $E_\mathrm{ij}$ (as in Eq.\ \ref{eq:4}), and a residual term for the three-particle interaction, $E_\mathrm{ijk}$.

\begin{equation}
 \label{eq:6}
E_\mathrm{interaction} = \Sigma_{\forall i} E_\mathrm{i} +  \Sigma_{\forall i<j} E_\mathrm{ij}(d_{H_i-H_j}) + E_\mathrm{ijk}
\end{equation}

It has been suggested in the past that the interaction between a cluster of hydrogen atoms in a metal can be expressed as a sum of single-particle and pairwise contributions \cite{haoHao2009Probing2009}. If this were true for hydrogen atoms in niobium, then the three-particle term would be zero in Eq.\ \ref{eq:6}, leading to Eq.\ \ref{eq:7}. 

\begin{equation}
 \label{eq:7}
E_\mathrm{interaction} = \Sigma_{\forall i} E_\mathrm{i} +  \Sigma_{\forall i<j} E_\mathrm{ij}(d_{H_i-H_j})
\end{equation}

We consider a $3\times 3\times 3$ niobium supercell with hydrogen atoms in different tetrahedral sites and hand-picked nine such structures with various interatomic distances between the hydrogen atoms. Structures $\sigma_1$, $\sigma_2$, $\sigma_3$, $\sigma_4$, $\sigma_5$ have three hydrogen atoms and therefore a hydrogen to metal ratio of $\frac{3}{54}\ or \sim 0.05$. Structures $\sigma_6$, $\sigma_7$, $\sigma_8$, $\sigma_9$ have twenty-seven hydrogen atoms and therefore a hydrogen to metal ratio of $\frac{27}{54} = 0.5$. We calculated the energies of the nine structures using DFT, allowing for the ion positions, cell shape, and the cell volume to be relaxed. \\

\begin{figure}[H]
\centering
 \includegraphics[width=\textwidth]{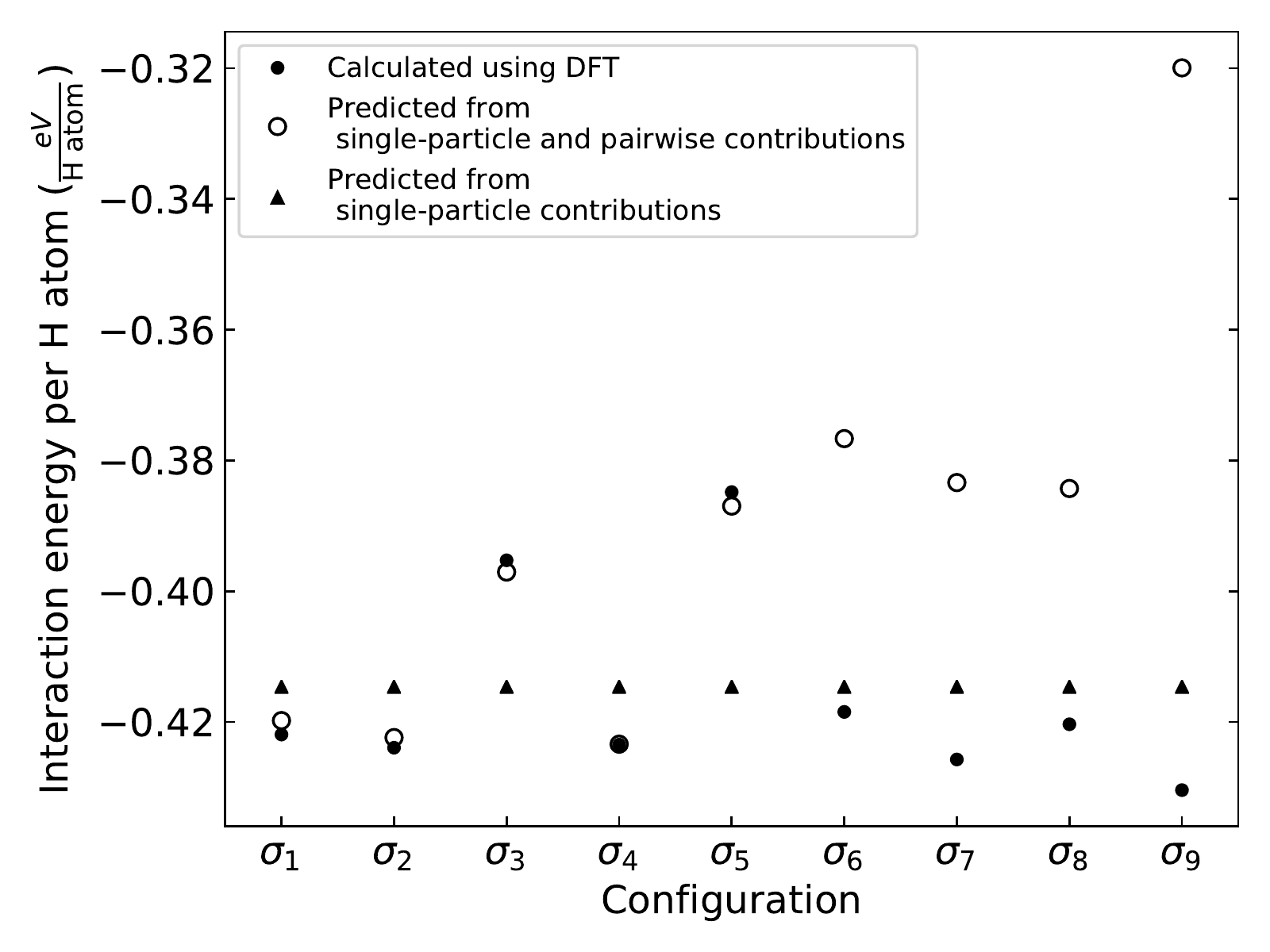}
 \caption{Interaction energies per hydrogen atom of the nine different structures of niobium-hydrogen systems. Structures $\sigma_1$, $\sigma_2$, $\sigma_3$, $\sigma_4$, $\sigma_5$ have $\frac{H}{M} \sim 0.05$. Structures $\sigma_6$, $\sigma_7$, $\sigma_8$, $\sigma_9$ have $\frac{H}{M} = 0.5$.}
 \label{fig:5}
\end{figure}

For the nine structures considered, we calculated the interaction energy according to Eq.\ \ref{eq:5}. We also calculated the interaction energy based on single-particle and pairwise contributions, as in Eq.\ \ref{eq:7}. We normalize these interaction energies per hydrogen atom and compare them in Fig.\ \ref{fig:5}. From Fig.\ \ref{fig:5}, we make the following inferences. For the five structures ($\sigma_1$, $\sigma_2$, $\sigma_3$, $\sigma_4$, $\sigma_5$) with $\frac{H}{M} \sim 0.05$, the real interaction energies calculated via Eq.\ \ref{eq:5} are pretty close to the interaction energies calculated from pairwise contributions via Eq.\ \ref{eq:7}, indicating that many-body characteristics are not significant in these structures. However, for all four structures ($\sigma_6$, $\sigma_7$, $\sigma_8$, $\sigma_9$) with $\frac{H}{M} = 0.5$, the real interaction energies calculated differ from the interaction energies calculated from pairwise contributions significantly. The offsets in prediction show that it is incorrect to think of the interaction energy as the sum of non-interacting binding energies and pairwise interaction energies for these structures. The results can be tied back to the shape of the enthalpy curve in Fig.\ 1. Based on Fukai's argument, the nonlinear variation of $\Delta H$ ($\frac{eV}{H atom}$) with hydrogen concentration indicates that the extensive enthalpy of dissolution ($eV$) is of at least of third order in hydrogen concentration, and in turn that the interactions between hydrogen atoms are not strictly pairwise, which was our conclusion for the structures having $\frac{H}{M} = 0.5$. However, at lower concentrations, the enthalpy curve is approximately linear in hydrogen concentration and thus, we don't expect many-body effects to be significant as is observed in the structures with $\frac{H}{M} \sim 0.05$, where a pairwise treatment of the interactions appears to be sufficient.  

\section{Conclusions}

We studied the interactions of interstitial hydrogen atoms in niobium by using DFT calculations applied to different systems such as single-hydrogen, double-hydrogen, and triple-hydrogen systems. Through these calculations, for the first time, we can understand these interactions from first principles and validate inferences from past experiments. These calculations reveal that there is an indirect image interaction between interstitial hydrogen atoms that is attractive and a direct interaction between interstitial hydrogen atoms that is repulsive. We also showed that these interactions have many-body characteristics. \\

We have shown that DFT is capable of resolving these complex issues and thus can offer further insights into the energy of different states. This understanding is necessary to sample thermodynamic ensembles and predict how hydrogen dissolution enthalpy varies with concentration and temperature.  The statistical mechanics formulation of this problem and its application to the prediction of absorption isotherms remains to be developed. 

\section{Acknowledgments}

We gratefully acknowledge the financial support from the Center for Negative Carbon Emissions at Arizona State University. We also acknowledge Research Computing at Arizona State University for providing high performance computing resources that have contributed to the research results reported within this work.

\section{Data Availability}

The authors confirm that the data required to reproduce the findings of this study are available within the article and can be reproduced by density functional theory calculations.

\bibliography{References}

\end{document}